\documentclass[10pt]{elsarticle} %3p - preview - review

\usepackage{geometry}
\usepackage{xcolor,colortbl} %testo colorato
\usepackage{hyperref}

\usepackage{multirow}
\usepackage{colortbl}
\usepackage{hhline}

\begin{document}

	\begin{center}
		\noindent{\Huge \textbf{A review on turbulent and vortical
		\\\vspace{6pt} flow analyses via complex networks}}
	\end{center}

	%% authors:
	\begin{center}
		{\Large G. Iacobello$^{1,}$\footnote[2]{Corresponding author: \textit{giovanni.iacobello\@ polito.it}\\Article published in Physica A, doi:\href{https://doi.org/10.1016/j.physa.2020.125476}{10.1016/j.physa.2020.125476}\\\textcopyright 2020. This manuscript version is made available under the CC-BY-NC-ND 4.0 license \href{http://creativecommons.org/licenses/by-nc-nd/4.0/}{(URL)}}, L. Ridolfi$^2$, S. Scarsoglio$^1$}
	\end{center}

	$^1$Department of Mechanical and Aerospace Engineering, Politecnico di Torino, Turin, 10129, Italy
	\\$^2$Department of Environmental, Land and Infrastructure Engineering, Politecnico di Torino, 10129 Turin, Italy

\vspace{1cm}
			\textbf{Abstract.}
			\\Turbulent and vortical flows are ubiquitous and their characterization is crucial for the understanding of several natural and industrial processes. Among different techniques to study spatio-temporal flow fields, complex networks represent a recent and promising tool to deal with the large amount of data on turbulent flows and shed light on their physical mechanisms. The aim of this review is to bring together the main findings achieved so far from the application of network-based techniques to study turbulent and vortical flows. A critical discussion on the potentialities and limitations of the network approach is provided, thus giving an ordered portray of the current diversified literature. The present review can boost future network-based research on turbulent and vortical flows, promoting the establishment of complex networks as a widespread tool for turbulence analysis.

\vspace{4pt}

	\section{Introduction}

		Fluid flows usually appear in nature and industrial applications as a turbulent motion, characterized by strong vortical structures and a chaotic behavior. Understanding turbulence (that is inherently vortical) and non-turbulent vortical flows (such as low Reynolds number wakes), hence, is crucial and has a relevant technological, environmental and economic impact. So far, several numerical and experimental techniques as well as theoretical models have been developed to unravel the intrinsic difficulties of dealing with turbulent and vortical flows~\cite{pope2001turbulent, wu2015vortical}. However, numerous issues -- such as coherent structure characterization, flow control, high Reynolds number scaling or emergence of spatio-temporal patterns -- are still open and subject of intense research~\cite{klewicki2010reynolds, marusic2010wall, sujith2020complex}.

	The high dimensionality and nonlinearity of fluid flows put strong methodological challenges to extract informative features from flow data, especially in case of turbulence. Novel methodological approaches are then continuously required to understand turbulent flows~\cite{pollard2017whither}. Several techniques have then been employed in fluid flow analysis so far, that have strongly relied on statistics (e.g., probability functions, structure functions, principal component decomposition~\cite{pope2001turbulent}) and analytical formulations (e.g., Fourier transform~\cite{pope2001turbulent}), as well as insights from dynamical systems theory (e.g., finite-time Lyapunov exponent~\cite{haller2001distinguished}). Each methodology has provided a different perspective on the flow dynamics; a very relevant example is the study in the frequency domain by Fourier analysis. However, facing open issues in turbulence still requires the development of different tools, which can overcome the limitations of traditional techniques or shed a different light on the complexity of turbulent flows.

		Novel approaches such as machine learning and network science, fostered by the increase in the computational capabilities and the availability of very large datasets, have come to light in recent times for studying fluid flows. Machine learning has gained well-grounded roots in several areas of fluid dynamics, and nowadays it is adopted as a tool for flow modeling, control and optimization strategies (e.g., see~\cite{brunton2020machine} for a comprehensive review). Compared with the wide diffusion of machine learning in fluid mechanics, the application of network science turbulent and vortical flow systems has a more recent development and less extended borders. While complex networks theory has been largely adopted in several research areas -- such as social networks, medicine, climate and economy, among others~\cite{costa2011analyzing} -- the first works trying to combine turbulence and complex networks date back to just a decade ago and only in the last few years a growing number of works has appeared, giving evidence of the rising interest from the turbulence scientific community. Although the adoption of a complex network viewpoint of turbulence is recent, different network methodologies have already been employed (in some cases  developed \textit{ad hoc}) to study turbulent and vortical flows under a variety of different flow configurations, producing so far a diversified collection of applications.
			
			The aim of this work is to gather the up-to-date literature review on network-based analysis of turbulent and vortical flows in an organized and critical view, which can be exploited as a reference point for future research on the topic. In fact, thanks to the preliminary attempts in building bridges between network science and turbulent and vortical flows, valuable insights have already emerged. With this review, therefore, we intend to highlight the main findings achieved so far, and provide a critical discussion on the potentialities and limitations of the network approach, thus trying to give an ordered portray of the current literature. Within the general context of fluid mechanics, a short review has been recently proposed on the application of network tools for climate and geophysical flows~\cite{donner2017focusissue}. In fact, climate and geophysical flows have been widely investigated so far by means of several network-based techniques~\cite{nocke2015visual, dijkstra2019networks}, also involving multi-scale or multi-layer network formulations~\cite{agarwal2019network, ying2020rossby}. Compared to~\cite{donner2017focusissue}, the present work comprises a different spectrum of topics involving turbulent and vortical flows, and applications are reviewed according to the flow configuration rather than to a method-wise fashion.
			
			The paper is organized as follows. Section~\ref{sec:methods} briefly summarizes the network-based methodologies typically used for turbulent flows analysis. The core of the review including the main results is in Section~\ref{sec:applications}. Section~\ref{sec:concl} draws some concluding remarks and future perspectives, while \ref{app:nets} provides a synthetic overview about complex networks and their metrics.

	\section{Mapping spatio-temporal flow data in complex networks}\label{sec:methods}

 		Methodologies developed to map fluid flows in complex networks can be organized in three main categories: (i) analysis of time-series; (ii) spatial network analysis; (iii) Lagrangian network analysis. The three methodologies are briefly outlined, highlighting the main features and drawbacks of each approach.

		\subsection{Network methods for time-series analysis}\label{subsec:timeseries}
			Three main classes of methodologies have been proposed so far to map (univariate or multivariate) time-series into networks~\cite{zou2019complex}: \textit{proximity networks}, \textit{transition networks} and \textit{visibility graphs}.
			
			\begin{figure}[t]
				\centering
				\includegraphics[width=.95\textwidth]{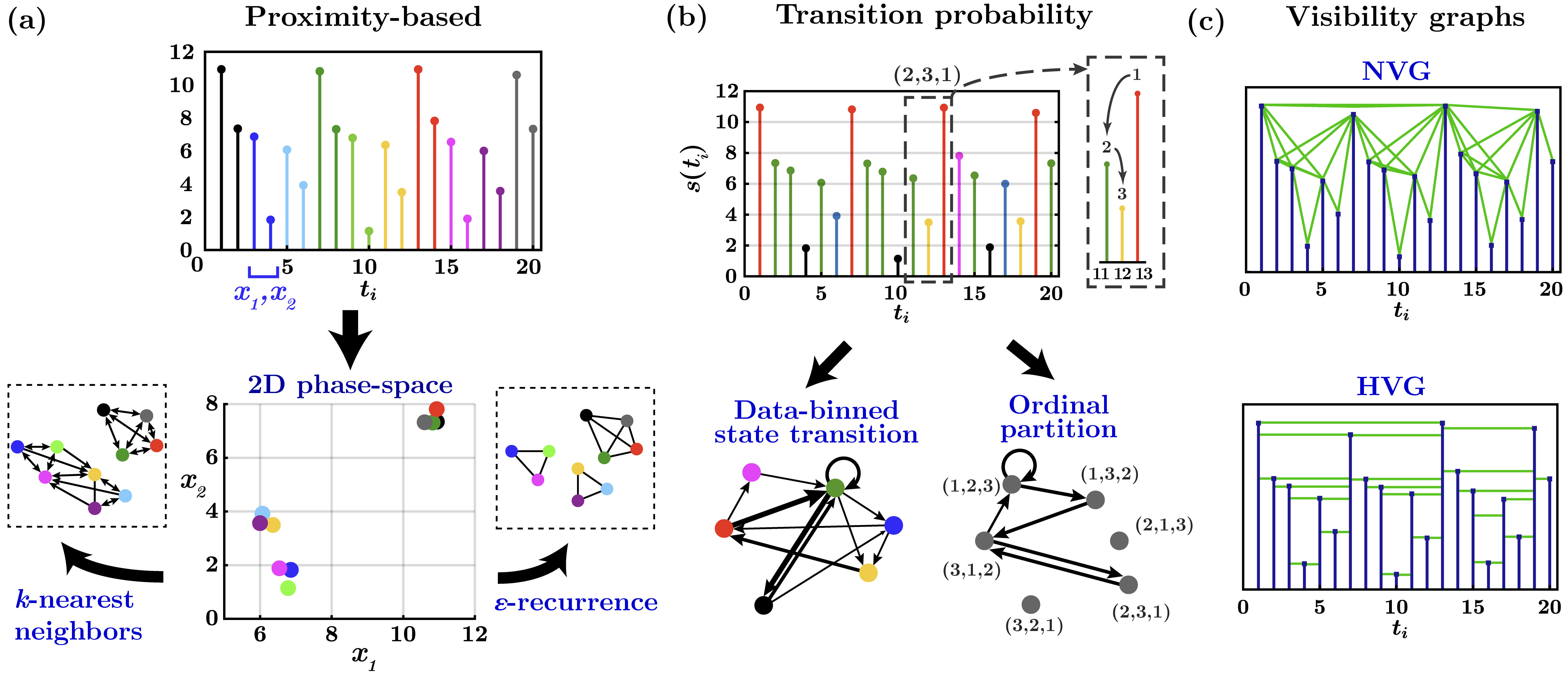}
				\caption{(a) Example of recurrence network building, in which a signal is divided in ten state vectors with dimension equal to two. The corresponding phase-space and a $k$-nearest neighbors ($k$-nn) network and an $\epsilon$-based network (with $k=3$ and $\epsilon=1$) are also shown. (b) Examples of transition networks. (Left), six states (i.e., nodes, highlighted with different colors) are obtained uniformly sampling the series in the range $\left(0,12\right)$; (Right) ordinal partition network with nodes equal to groups of three consecutive data (the ordinal descending sequence is indicated within brackets). The link thickness is proportional to the transition probability between two consecutive nodes. (c) Example of natural (NVG) and horizontal (HVG) visibility graphs, where links are depicted as green lines and data are shown as blue bars.}\label{fig:mets_ts}
			\end{figure}

			\subsubsection{Proximity networks}\label{subsubsec:proximity_nets}
				In proximity networks, nodes are associated with groups of data referred to as \textit{state vectors} and links are activated on the basis of a proximity in a suitable phase-space. Two approaches are usually followed, namely, recurrence networks~\cite{marwan2009complex} and cycle networks~\cite{zhang2006complex}. In recurrence networks, states (i.e., nodes) are embedded in a phase-space with size equal to the state length (i.e., the \textit{embedding dimension}). For instance, in \figurename~\ref{fig:mets_ts}a, states are consecutive pairs of data (depicted with different colors); thus the phase-space has dimension equal to two. Links are enabled if two nodes are sufficiently close in the phase-space according to a specific metric, such as the Euclidean distance. The general idea, therefore, is to interpret the recurrence matrix employed for recurrence plot analysis as a network adjacency matrix~\cite{marwan2009complex}. Among other proposed linking criteria, the $k$-nearest neighbors ($k$-nn) and $\epsilon$-recurrence approaches have been employed for studying vortical and turbulent flows. In $k$-nn networks, a node is linked to the closest $k$ nodes in the phase-space, so that each node degree (see~\ref{app:nets}) is fixed (in \figurename~\ref{fig:mets_ts}a, $k=3$). In $\epsilon$-recurrence networks, instead, links are established if the Euclidean distance (defined in the phase-space) is less than a threshold $\epsilon$ (in \figurename~\ref{fig:mets_ts}a, $\epsilon=1$). 
				
				In both $k$-nn and $\epsilon$-recurrence networks, a set of (arbitrary) parameters needs to be properly tuned. The choice of the embedding dimension value (e.g., equal to two nodes in \figurename~\ref{fig:mets_ts}a) is common to all recurrence network approaches, and relies on characteristic times associated with the system under study (e.g., characteristic periods). Additionally, states (i.e., nodes) are usually defined as time-delayed vectors -- namely intervals of the signal with a temporal spacing larger than the sampling time-step~\cite{marwan2009complex} -- so that a time-delay needs also to be defined (e.g., in \figurename~\ref{fig:mets_ts}a the time-delay is equal to the time-step since nodes are defined by consecutive data). In $\epsilon$-recurrence networks, the choice of the threshold $\epsilon$ is non-trivial and is typically set in order to fix the network edge density~\cite{takagi2017nonlinear, kobayashi2018nonlinear}, to make the network be fully connected~\cite{godavarthi2017recurrence, murugesan2019complex}, as well as by relying on percolation analysis that produces a single giant component~\cite{gotoda2017characterization}. In applications of $k$-nn to turbulent flows (see~\cite{charakopoulos2014application, okuno2015dynamics}), a value $k=4$ has been used (as suggested in~\cite{xu2008superfamily}), since small values of $k$ in the range $\left[3-8\right]$ are usually able to capture dynamical differences in the system.
				
				Cycle networks are employed to study periodic-like time-series~\cite{zhang2006complex}. Due to the pseudo-periodicity in the signal, nodes are identified with temporal intervals of the signal, called cycles, between two consecutive local maxima (or minima). Links can be established if the intervals of series representing cycles are close in a phase-space or are sufficiently correlated.

			\subsubsection{Transition networks}\label{subsubsec:trans_nets}
			
			Transition matrices (also called probability matrices) are the representation of a Markov chain in the context of stochastic modeling. While studies based on probability matrices have largely been employed in several research fields including fluid mechanics (e.g., see~\cite{Kaiser2014Cluster, nair2019cluster}), transition networks analysis has recently emerged as a tool to explicitly exploit the topological properties of transition matrices~\cite{zou2019complex}. The key feature of transition networks is that a link between two nodes is weighted according to the (transition) probability that one node directly follows in time the other node~\cite{zou2019complex}, so that transition networks are usually directed graphs~\cite{newman2018networks}. Two examples of transition networks are shown in \figurename~\ref{fig:mets_ts}b for a given signal, $s(t_i)$. In the top panel of \figurename~\ref{fig:mets_ts}b the values of $s(t_i)$ are divided in six equally-spaced states (i.e., nodes, which are highlighted with different colors) from $0$ to $12$. The corresponding network (bottom left panel) comprises six nodes and links are established if there is a direct transition in time from one state to another state (e.g., there is no transition from values in red to values in black). 
				\\The second example is an \textit{ordinal-partition transition-network}, as proposed in~\cite{mccullough2015time}. A state corresponds to a group of series-values sampled at a given temporal delay (e.g., in \figurename~\ref{fig:mets_ts}b states have length three with time-delay equal to the signal time-step, namely $\Delta t_i=1$), and an ordinal (ascending or descending) sequence is assigned to each state according to the position of the series-values in the sequence. An example of state with length 3, $t_i=\lbrace 11,12,13\rbrace$, is highlighted in the top panel of \figurename~\ref{fig:mets_ts}b (see dashed grey boxes) together with its symbolic descending sequence. The descending inequality for $t_i=\lbrace 11,12,13\rbrace$ is $s(13)>s(11)>s(12)$ and the corresponding descending sequence is $(2,3,1)$, because $t_i=13$ occupies the first position in the inequality (i.e., the $1$ in the sequence indicates which is the maximum value in the state), while $t_i=11$ and $t_i=12$ occupy the second and third positions, respectively. Each integer in a sequence, therefore, refers to the position of each datum $s(t_i)$ in the descending or ascending inequality. Any possible sequence corresponds to a network node (in \figurename~\ref{fig:mets_ts}b each state has three values so the number of nodes is $3!=6$) and links are established according to the temporal transition between consecutive sequences. The bottom right panel of \figurename~\ref{fig:mets_ts}b shows the ordinal transition network for $s(t_i)$; nodes corresponding to sequences $(3,2,1)$ and $(2,1,3)$ do not appear in the signal so they are disconnected nodes.

			\subsubsection{Visibility graphs}\label{subsubsec:vis_graphs}
			
				The third large class of networks for time-series analysis is represented by visibility graphs, in which nodes represent single values of the series. Two main variants have been proposed~\cite{zou2019complex}, as shown in \figurename~\ref{fig:mets_ts}c. The natural visibility graph (NVG) exploits a convexity criterion~\cite{lacasa2008time}, so that a link between two nodes is activated if the straight line connecting the two data points lies above the other in-between data. E.g., in \figurename~\ref{fig:mets_ts}c, node $3$ (i.e., at $t_i=3$) is connected to node $5$ (i.e., at $t_i=5$) since node $4$ is below the straight line (highlighted in green) between nodes $3$ and $5$. A simplified version is the horizontal visibility graph (HVG) that is an ordering criterion~\cite{luque2009horizontal}, namely two nodes are linked if the series values of the two nodes are both higher than the value of any other in-between datum. Geometrically, two nodes are linked in a HVG network if the horizontal line connecting the two nodes lies above the other in-between data (see green lines in \figurename~\ref{fig:mets_ts}c). Thanks to its simplicity, analytical results have been given for HVGs from random signals~\cite{luque2009horizontal}. Both NVGs and HVGs are invariant under affine transformations of the series, i.e. under rescaling and translations of both horizontal and vertical axes. This implies that two time-series with a similar temporal structure but different mean and standard deviation values are mapped into the same visibility network~\cite{iacobello2019experimental}. Differently from proximity and transition networks (that usually require the choice of at least a parameter), visibility graphs are parameter-free.

	\begin{figure}[t]
				\centering
				\includegraphics[width=\textwidth]{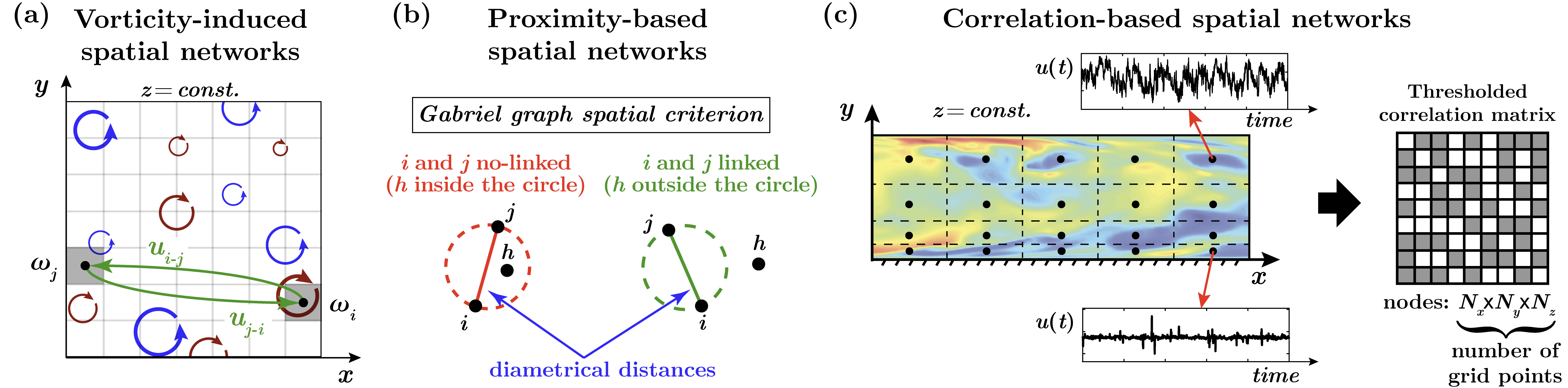}
				\caption{(a) Example of a 2D vortical field represented via rotating arrows with intensity $\omega$ (highlighted with a different line thickness); blue and red arrows indicate positive and negative vorticity, respectively. The induced velocity, $u_{i-j}$ and $u_{j-i}$, between two spatial locations, $i$ and $j$, serve as link weights and are illustrated via green arrows. (b) Example of proximity-based network building via the Gabriel graph approach. (c) Example of correlation-based network building. A discretization of the domain in cells (represented by grid points in black, which correspond to network nodes) is performed. Velocity time-series in a turbulent channel flow (a snapshot of the streamwise velocity fluctuations is shown as a 2D colored section) are extracted at each grid point and exploited to build a correlation coefficient matrix, whose entries are thresholded to obtain an adjacency matrix (shown as a checkerboard  plot).} \label{fig:mets_eul}
			\end{figure}

		\subsection{Network-based analysis in the Eulerian viewpoint}\label{subsec:Eul_meth}

			Spatial networks refer to graphs in which nodes occupy specific (fixed) locations in space. Within the Eulerian viewpoint of vortical and turbulent flows, three main approaches (see \figurename~\ref{fig:mets_eul}) have been proposed: (i) vorticity-induced velocity-interactions, (ii) proximity-based, and (iii) correlation-based link activation. In the former case (e.g., see \figurename~\ref{fig:mets_eul}a), a node is associated to a vortical fluid element, and links between pairs of nodes (depicted as green arrows in \figurename~\ref{fig:mets_eul}a) are quantified through the amount of induced velocity (dictated by the Biot-Savart law) between them~\cite{nair2015network, taira2016network}. Accordingly, vorticity-induced spatial networks are typically link-weighted directed graphs; undirected networks can also be obtained by combining the information on the incoming and outgoing links.
			\\In proximity-based spatial networks (as for time-series) links are simply activated if nodes occupy nearby locations in the physical space. For instance, \figurename~\ref{fig:mets_eul}b shows a sketch of a proximity network based on the Gabriel graph criterion~\cite{krueger2019quantitative}. According to this criterion, two distinct nodes, $i$ and $j$, are linked (as shown in the green case on the right of \figurename~\ref{fig:mets_eul}b) if any other node, $h$, lies outside the circle (illustrated as dashed lines in \figurename~\ref{fig:mets_eul}b) whose diameter is the distance between $i$ and $j$.			
			\\Finally, one of the most common way to build spatial networks is via correlation of physical quantities (e.g., the velocity components) in different spatial locations (e.g., see \figurename~\ref{fig:mets_eul}c in which a correlation network is built for a turbulent channel flow)~\cite{aste2006dynamical, donges2009complex, zhou2015teleconnection, tupikina2016correlation}. In this approach, correlation coefficients are typically filtered via a threshold that retains the highest values (see the checkerboard plot in \figurename~\ref{fig:mets_eul}c), so that only the strongest (positive and-or negative) values are retained.

		\subsection{Complex networks from Lagrangian particle trajectories}\label{subsec:lagrang_meth}
		
\begin{figure}[t]
				\centering
				\includegraphics[width=\textwidth]{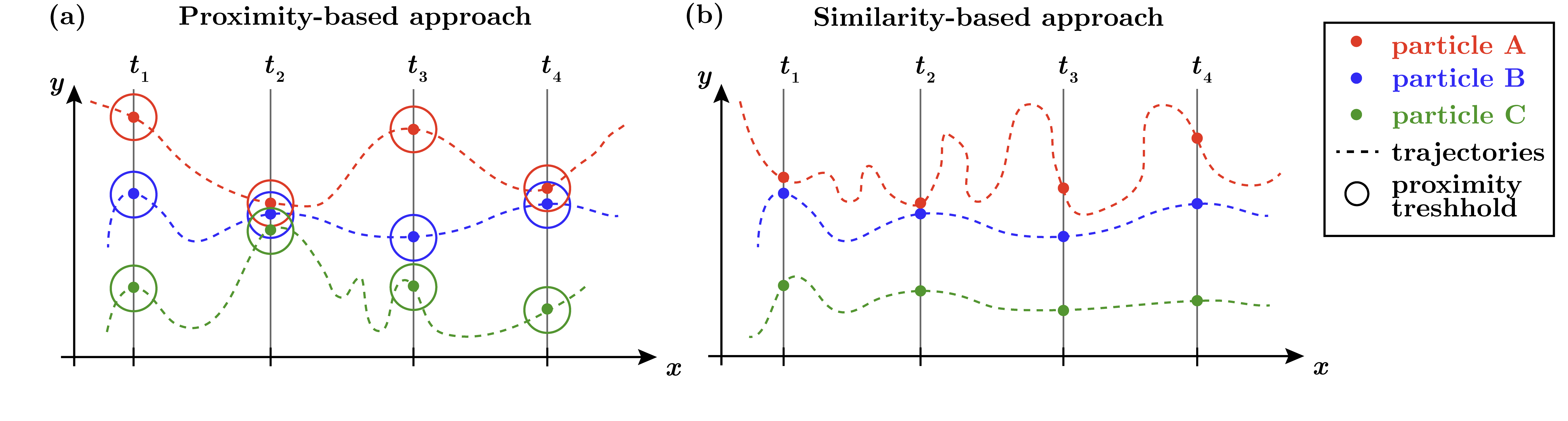}
				\caption{Sketch of three particles (colored dots) and 2D projection of their trajectories in the physical space (colored dashed lines). Circles indicate the proximity-threshold distance for link activation for proximity-based Lagrangian networks.} \label{fig:mets_lagr}
			\end{figure}

			The third way to build networks from turbulent data exploits a Lagrangian viewpoint in which the position of a set of (actively or passively driven) entities is tracked over time. In this framework, particles (or, equivalently, their trajectory) or fluid elements as point vortices are typically associated to nodes of the network. Three main categories of Lagrangian networks have been proposed: \textit{vorticity-induced networks}~\cite{meena2018network}, \textit{proximity-based} and \textit{similarity-based} networks (see \figurename~\ref{fig:mets_lagr}). Vorticity-induced Lagrangian networks are equivalent to the Eulerian counterpart (see \figurename~\ref{fig:mets_eul}a) and are obtained by tracking vortical elements in the domain~\cite{meena2018network}. In proximity-based networks, a link between a particle pair is established if the two particles come sufficiently close in space during their motion~\cite{hadjighasem2016spectral}. E.g., in \figurename~\ref{fig:mets_lagr}a, particles A and B are sufficiently close at times $t_2$ and $t_4$, while particles B and C only at time $t_2$. The proximity criterion between particles can be checked within a given time interval (in \figurename~\ref{fig:mets_lagr}a, particle B is connected to particles A and C at least once in the interval $t_1-t_4$)~\cite{padberg2017network, schneide2018probing}, or at each time instant (in \figurename~\ref{fig:mets_lagr}a, particle B is connected to both particles A and C only at time $t_2$)~\cite{iacobello2019lagrangian}. In the latter case, a temporal network is obtained, i.e., a network whose structure evolves in time. In similarity-based networks (\figurename~\ref{fig:mets_lagr}b), instead, two particles are linked if they follow trajectories with similar path shapes regardless of their physical proximity to one another. E.g., particles A and B in \figurename~\ref{fig:mets_lagr}b come close in space during their motion but follow different trajectories (A and B are not linked), while particles B and C follow very similar trajectories (B and C are linked). Therefore, the similarity-based criterion requires the definition of an adequate time window for the comparison of trajectories. An example of similarity criterion is represented by the standard deviation of the time-series of the distance between two trajectories, where low standard deviation values indicate similar trajectories~\cite{schlueter2017coherent, husic2019simultaneous}.
			
		Finally, particle trajectories can be used to build Lagrangian flow-networks, in which nodes represent fixed spatial locations while links are weighted according to the amount of particles that are transported between different locations~\cite{ser2015flow}. In this way, the resulting network can be interpreted as a transition network in which links quantify the probability that particles are transported from one node to another node in the network~\cite{ser2015flow}.

	\section{Network-based applications to turbulent and vortical flows}\label{sec:applications}
	
	This Section reports the main achievements obtained so far from the application of network tools to turbulent and vortical flows. The information on the methodologies of each reviewed work is outlined in \tablename~\ref{tab:methds}, according to the definitions of nodes and linking criterion (for methods details, refer to Section~\ref{sec:methods}).
	
	\begin{table}[t]
		\caption{Summary of the network methodologies employed in the reviewed literature. Acronyms legend: \textbf{NVG}, natural visibility graph; \textbf{HVG}, horizontal visibility graph; $k$-nn, $k$-nearest neighbors; \textbf{RBC}, Rayleigh-B{\'e}nard convection; \textbf{DDC}, double-diffusive convection; \textbf{WT}, wall turbulence; \textbf{POD}, proper orthogonal decomposition; \textbf{STD}, standard deviation.}\label{tab:methds}
		\scriptsize
 		\setlength{\tabcolsep}{2pt} %regola distanza testo da linee
 		\renewcommand{\arraystretch}{1.2}	
 		\vspace{2mm}
		\begin{tabular}{cc|c|c|c|c|c|}
			 %header
			 \multicolumn{1}{m{.09\textwidth}}{\multirow{2}{.1\textwidth}{}} & \multicolumn{1}{|m{.095\textwidth}|}{\multirow{1}{.095\textwidth}{\centering\textit{\textbf{Linking criterion}}}} & \multirow{1}{.11\textwidth}{\centering\textbf{Isotropic Turbulence}} & \multirow{1}{.1650\textwidth}{\centering\textbf{Jets and plumes}} & \multirow{1}{.145\textwidth}{\centering\textbf{Wakes and vortical flows}} & \multirow{1}{.175\textwidth}{\centering\textbf{Wall turbulence and convection}} & \multirow{1}{.145\textwidth}{\centering\textbf{Turbulent combustors}}\\
			   & \multicolumn{1}{|m{.095\textwidth}|}{} & & & & &\\ \hline
			  %			  
			  % time-series:
			  \multicolumn{1}{m{.09\textwidth}}{\multirow{2}{.1\textwidth}{\cellcolor[rgb]{0.675,0.306,0.059}}} & \multicolumn{1}{|m{.095\textwidth}|}{\cellcolor[rgb]{0.87,0.5,0.19}\centering \textbf{\textit{Proximity}}} &  \cellcolor[rgb]{0.87,0.5,0.19}  & \multicolumn{1}{m{.1650\textwidth}|}{\cellcolor[rgb]{0.87,0.5,0.19}\centering $\epsilon$\textbf{-recurrence}: \cite{takagi2017nonlinear, murugesan2019complex}; $k$\textbf{-nn}: \cite{charakopoulos2014application}} &  \cellcolor[rgb]{0.87,0.5,0.19}  & \cellcolor[rgb]{0.87,0.5,0.19}  & \multicolumn{1}{m{.145\textwidth}|}{\cellcolor[rgb]{0.87,0.5,0.19}$\epsilon$\textbf{-recurrence}:\cite{godavarthi2017recurrence, kobayashi2018nonlinear}; $k$\textbf{-nn}:\cite{okuno2015dynamics}; \textbf{Cycle networks}:\cite{okuno2015dynamics}}\\
			  \multirow{-3}{.09\textwidth}{\cellcolor[rgb]{0.675,0.306,0.059}\centering\textcolor{white}{\textbf{Signal network analysis}}} & \multicolumn{1}{|m{.095\textwidth}|}{\cellcolor[rgb]{0.957,0.69,0.518}\centering \textbf{\textit{Transition}}} & \cellcolor[rgb]{0.957,0.69,0.518}  & \multicolumn{1}{m{.1650\textwidth}|}{\cellcolor[rgb]{0.957,0.69,0.518}Nodes $\rightarrow$ data-binned signal values\,\cite{shirazi2009mapping}} & \cellcolor[rgb]{0.957,0.69,0.518}  & \multicolumn{1}{m{.175\textwidth}|}{\cellcolor[rgb]{0.957,0.69,0.518}\textbf{\textit{WT}}: Nodes $\rightarrow$ points in phase-space\,\cite{schmid2018description}} &  \multicolumn{1}{m{.145\textwidth}|}{\cellcolor[rgb]{0.957,0.69,0.518}\centering \textbf{Ordinal partition} \cite{kobayashi2019early, kobayashi2018nonlinear, kobayashi2019spatiotemporal, aoki2020dynamic}}\\  
			  \cellcolor[rgb]{0.675,0.306,0.059} & \multicolumn{1}{|m{.095\textwidth}|}{\cellcolor[rgb]{0.973,0.796,0.678}\centering \textbf{\textit{Visibility}}} & \cellcolor[rgb]{0.973,0.796,0.678}  & \multicolumn{1}{m{.1650\textwidth}|}{\cellcolor[rgb]{0.973,0.796,0.678}\centering \textbf{NVG}:\cite{charakopoulos2014application, murugesan2019complex, iacobello2018complex, iacobello2019experimental}; \textbf{HVG}: \cite{manshour2015fully, takagi2018dynamic, kobayashi2019spatiotemporal}; \textbf{Spatial VG}:\cite{Tokami2020spatiotemporal}} & \multicolumn{1}{m{.145\textwidth}|}{\cellcolor[rgb]{0.973,0.796,0.678}\centering\textbf{HVG}:\,\cite{tao2019modified, wu2020evolution}} & \multicolumn{1}{m{.175\textwidth}|}{\cellcolor[rgb]{0.973,0.796,0.678}\centering\textbf{\textit{WT}}: \textbf{NVG}\,\cite{liu2010statistical, iacobello2018visibility}; \textbf{\textit{DDC}}: \textbf{HVG}\,\cite{kondo2018chaotic}} & \multicolumn{1}{m{.145\textwidth}|}{\cellcolor[rgb]{0.973,0.796,0.678}\centering \textbf{NVG}:\,\cite{murugesan2015combustion, murugesan2016detecting}; \textbf{HVG}: \cite{guan2020intermittency}; \textbf{Spatial VG}:\,\cite{singh2017network}}\\ \hline	
			  %
			  % spatial:
			  \multicolumn{1}{m{.09\textwidth}}{\cellcolor[rgb]{0.33,0.51,0.21}} & \multicolumn{1}{|m{.095\textwidth}|}{\cellcolor[rgb]{0.439,0.678,0.278}\centering \textbf{\textit{Vorticity-induced}}} & \multicolumn{1}{m{.11\textwidth}|}{\cellcolor[rgb]{0.439,0.678,0.278}\centering Nodes $\rightarrow$ points in a 2D grid \cite{taira2016network, bai2019randomized}} & \multicolumn{1}{m{.1650\textwidth}|}{\cellcolor[rgb]{0.439,0.678,0.278}\centering Nodes $\rightarrow$ points in a 2D grid\,\cite{takagi2018synchronization, takagi2018effect, kobayashi2019spatiotemporal}} & \multicolumn{1}{m{.145\textwidth}|}{\cellcolor[rgb]{0.439,0.678,0.278}\centering Nodes $\rightarrow$ points in a 2D grid\,\cite{meena2018network, bai2019randomized}} &  \cellcolor[rgb]{0.439,0.678,0.278}  & \multicolumn{1}{m{.145\textwidth}|}{\cellcolor[rgb]{0.439,0.678,0.278}\centering Nodes $\rightarrow$ points in a 2D grid \cite{murayama2018characterization, hashimoto2019spatiotemporal}}\\
			  \multirow{-2}{.09\textwidth}{\cellcolor[rgb]{0.33,0.51,0.21}\centering\textcolor{white}{\textbf{Spatial networks}}} & \multicolumn{1}{|m{.095\textwidth}|}{\cellcolor[rgb]{0.588,0.780,0.463}\centering \textbf{\textit{Proximity}}} &  \cellcolor[rgb]{0.588,0.780,0.463}  & \cellcolor[rgb]{0.588,0.780,0.463}  & \multicolumn{1}{m{.145\textwidth}|}{\cellcolor[rgb]{0.588,0.780,0.463}Nodes $\rightarrow$ critical points of the velocity field, Gabriel graph criterion\,\cite{krueger2019quantitative}} &  \cellcolor[rgb]{0.588,0.780,0.463}  & \multicolumn{1}{m{.145\textwidth}|}{\cellcolor[rgb]{0.588,0.780,0.463}Nodes $\rightarrow$ points in a 2D grid, time-varying graph\,\cite{krishnan2019emergence}}\\
			  \cellcolor[rgb]{0.33,0.51,0.21} & \multicolumn{1}{|m{.095\textwidth}|}{\cellcolor[rgb]{0.776,0.878,0.706}\centering \textbf{\textit{Correlation}}} & \multicolumn{1}{m{.11\textwidth}|}{\cellcolor[rgb]{0.776,0.878,0.706}\centering Nodes $\rightarrow$ points in a 3D grid \cite{scarsoglio2016complex}} & \cellcolor[rgb]{0.776,0.878,0.706}  &  \cellcolor[rgb]{0.776,0.878,0.706}  & \multicolumn{1}{m{.175\textwidth}|}{\cellcolor[rgb]{0.776,0.878,0.706}\textbf{\textit{WT}}: Nodes $\rightarrow$ volumes of a 3D non-uniformly spaced grid\,\cite{iacobello2018spatial}} & \multicolumn{1}{m{.145\textwidth}|}{\cellcolor[rgb]{0.776,0.878,0.706}\centering Nodes $\rightarrow$ 2D grid points\,\cite{unni2018emergence, krishnan2019mitigation}}\\ \hline	
			  %
			  %lagrangian:
			  \multicolumn{1}{m{.09\textwidth}}{\cellcolor[rgb]{0.184,0.459,0.71}} &\multicolumn{1}{|m{.095\textwidth}|}{\cellcolor[rgb]{0.36,0.61,0.84}\centering \textbf{\textit{Vorticity-induced}}} &  \cellcolor[rgb]{0.36,0.61,0.84}  &  \cellcolor[rgb]{0.36,0.61,0.84}  & \multicolumn{1}{m{.145\textwidth}|}{\cellcolor[rgb]{0.36,0.61,0.84}\centering Nodes $\rightarrow$ point vortices\,\cite{meena2018network, nair2015network}} & \cellcolor[rgb]{0.36,0.61,0.84}  & \cellcolor[rgb]{0.36,0.61,0.84}  \\ 
			  \cellcolor[rgb]{0.184,0.459,0.71} & \multicolumn{1}{|m{.095\textwidth}|}{\cellcolor[rgb]{0.61,0.761,0.902}\centering \textbf{\textit{Proximity}}} &  \cellcolor[rgb]{0.61,0.761,0.902}  &  \cellcolor[rgb]{0.61,0.761,0.902}  & \multicolumn{1}{m{.145\textwidth}|}{\cellcolor[rgb]{0.61,0.761,0.902}\centering Particle proximity occurs at least once\,\cite{padberg2017network}} &  \multicolumn{1}{m{.175\textwidth}|}{\cellcolor[rgb]{0.61,0.761,0.902}\textbf{\textit{WT}}: Particle proximity checked at each time $\rightarrow$ time-varying network\,\cite{iacobello2019lagrangian}; \textbf{\textit{RBC}}: Particle proximity occurs at least once\,\cite{schneide2018probing}} & \cellcolor[rgb]{0.61,0.761,0.902}  \\ 
			  \multirow{3}{.09\textwidth}{\cellcolor[rgb]{0.184,0.459,0.71}\centering\textcolor{white}{\textbf{Lagrangian networks}}} & \multicolumn{1}{|m{.095\textwidth}|}{\cellcolor[rgb]{0.74,0.843,0.933}\centering \textbf{\textit{Similarity}}} & \cellcolor[rgb]{0.74,0.843,0.933}  & \cellcolor[rgb]{0.74,0.843,0.933}  & \multicolumn{1}{m{.145\textwidth}|}{\cellcolor[rgb]{0.74,0.843,0.933}Link weights $\rightarrow$ STD of the distance between trajectories (nodes)\,\cite{schlueter2017coherent, schlueter2017identification, husic2019simultaneous}} &  \cellcolor[rgb]{0.74,0.843,0.933}  & \cellcolor[rgb]{0.74,0.843,0.933}  \\
			  \cellcolor[rgb]{0.184,0.459,0.71} & \multicolumn{1}{|m{.095\textwidth}|}{\cellcolor[rgb]{0.686,0.912,0.953}\centering \textbf{\textit{Flow-networks}}} & \cellcolor[rgb]{0.686,0.912,0.953}  & \cellcolor[rgb]{0.686,0.912,0.953}  & \cellcolor[rgb]{0.686,0.912,0.953} &  \multicolumn{1}{m{.175\textwidth}|}{\cellcolor[rgb]{0.686,0.912,0.953} \textbf{\textit{WT}}: Nodes $\rightarrow$ wall-normal levels; time-varying links proportional to the number of particles flowing among levels\,\cite{perronewall}}  & \cellcolor[rgb]{0.686,0.912,0.953}  \\ \hline
			  %
			  %others
			  \multicolumn{1}{m{.09\textwidth}}{\cellcolor[rgb]{0.251,0.251,0.251}\centering\textcolor{white}{\textbf{Others}}} & \multicolumn{1}{|m{.095\textwidth}|}{\cellcolor[rgb]{0.627,0.627,0.627}\centering \textbf{\textit{Oscillatory networks}}} &  \cellcolor[rgb]{0.627,0.627,0.627}  &  \cellcolor[rgb]{0.627,0.627,0.627}  & \multicolumn{1}{m{.145\textwidth}|}{\cellcolor[rgb]{0.627,0.627,0.627}\centering Nodes $\rightarrow$ POD mode pairs\,\cite{nair2018networked}} &  \cellcolor[rgb]{0.627,0.627,0.627}  &  \cellcolor[rgb]{0.627,0.627,0.627} \\ 			  		 			 
			\hline	
		\end{tabular}	
	\end{table}

		\subsection{Isotropic turbulence}\label{subsec:isotrp}

			One of the simplest configurations for turbulence is obtained in case of isotropy~\cite{pope2001turbulent}. \textit{Taira et al.}~\cite{taira2016network} studied a 2D decaying isotropic turbulent flow in which the link intensity was quantified by means of vorticity-induced velocity (e.g., see \figurename~\ref{fig:mets_eul}a). By doing so, the strongest vortical elements in the flow were mapped in the network hubs (i.e., the most connected nodes). The authors showed that, when both large- and small-scale vortices are present, the vorticity-network displays a scale-free structure, namely a power-law distribution for the node strength (i.e., the node weighted degree~\cite{newman2018networks}, see~\ref{app:nets}). The decay in turbulence energy with time corresponds to a disappearance of the smallest vortices, so that only the largest vortical elements survive and the network scale-freeness is lost. Physically, the scale-free property of the vorticity-network indicates that the flow does not possess a characteristic vortical scale (i.e., vortical structures are self-similarly distributed in the flow), and that strong vortical structures (hubs) are rare. These two features have important practical implications, since hubs can be targeted to implement flow control strategies. In fact, \textit{Taira et al.}~\cite{taira2016network} evidenced that vorticity networks are resilient under random perturbations (i.e., random node removals) of the flow field, but not under hubs perturbation, namely the removal of the strongest vortical structures significantly affects the flow structure.

			Differently from the work in~\cite{taira2016network}, \textit{Scarsoglio et al.}~\cite{scarsoglio2016complex} proposed a correlation-threshold approach (see Section~\ref{subsec:Eul_meth}) to investigate a 3D (forced) homogeneous isotropic flow. By exploiting time-series of turbulent kinetic energy to calculate the correlation coefficient, the authors found out the appearance of spatial patterns of high degree nodes (i.e., network hubs) within the Taylor microscale, in spite of the homogeneity and isotropy of the flow. Most importantly, such patterns of hubs revealed to be associated to spatial regions that coherently evolve with similar vorticity values. Hence, the correlation-network based on the turbulent kinetic energy is able to infer the higher order spatial information of the vorticity field (which involves velocity gradients), i.e., coherent patterns over the temporal window considered. This capability of complex networks to identify flow critical regions (such as coherent eddies) in a texture of interconnections is a key feature that has been leveraged in several studies on turbulent and vortical flows, as it will be discussed in the next subsections.

		\subsection{Jets and Plumes}\label{subsec:jets}
			
			A common way in which turbulence appears in nature and industrial applications is represented by jets and plumes. These two flow configurations have been the subject of several network-based analyses, even including heat sources (such as a combustion process) or a flow interaction with external factors (e.g., obstacles, a background turbulent flow or an external forcing). 
			
			\textit{Shirazi et al.}~\cite{shirazi2009mapping} and \textit{Manshour et al.}~\cite{manshour2015fully} investigated linear (e.g., autocorrelation structure) and non-linear dependencies of turbulence signals via transition and horizontal visibility networks, respectively. With respect to surrogate random series, long-range correlation was highlighted in velocity signals of a free jet~\cite{shirazi2009mapping}, and linear (e.g., autocorrelation structure) and non-linear dependencies were detected in velocity and acceleration time-series of low-temperature jets~\cite{manshour2015fully}. Both network approaches were tested for different Reynolds numbers, showing the sensitivity of complex networks to the specific flow conditions. Furthermore, the technique proposed in~\cite{shirazi2009mapping} allows one to perform a signal reconstruction from the network topology, and the authors showed excellent agreement of PDF and structure functions between original and reconstructed signals.

		The characterization of turbulent vortical structures with different size in jets and plumes has been pursued both for numerical~\cite{kobayashi2019spatiotemporal} and experimental data~\cite{charakopoulos2014application, iacobello2018complex, iacobello2019experimental}, aiming to highlight how vortical structures affect the topology of networks built from time-series. A water-into-water heated jet was focused by \textit{Charakopoulos et al.}~\cite{charakopoulos2014application} who were able to identify different spatial regions in the jet flow through $k$-nn and HVG approaches, so that large/small values of the network metrics were related to the occurrence of small or large vortical structure in the flow. In a fashion similar to~\cite{charakopoulos2014application}, \textit{Kobayashi et al.}~\cite{kobayashi2019spatiotemporal} found significant variations of the metrics in HVGs and vorticity-induced spatial networks exploiting vorticity time-series from a coaxial jet in a turbulent combustor. In both works, the network analysis revealed that metric variations are associated to the presence of short-living small vortical structures along the jet axis~\cite{charakopoulos2014application, kobayashi2019spatiotemporal}. The interaction between a passive scalar plume and turbulent structures in a turbulent boundary layer, was studied by \textit{Iacobello et al.}~\cite{iacobello2018complex, iacobello2019experimental} in terms of meandering motion (associated with eddies larger than the plume size) and relative dispersion of the plume (associated with eddies smaller than the plume size). Important features as the occurrence, the relative intensity and the temporal collocation of extreme events (peaks and pits) were highlighted through the NVG analysis of concentration~\cite{iacobello2018complex} and turbulent transport time-series~\cite{iacobello2019experimental}. The network metrics advanced the knowledge about extreme events, e.g., discerning between peaks and outliers, thus enriching the analysis via common statistical tools as spectra and higher-order moments.

		Network-based analyses have also been carried out to evaluate the randomness of plume dynamic behaviors in presence of buoyancy effects and combustion processes. Turbulent fires displayed an increase in network randomness -- of $k$-nn and visibility networks -- by moving towards the far field of the plume~\cite{takagi2017nonlinear, takagi2018dynamic, Tokami2020spatiotemporal}, distinguishing between a large-scale vortex rings-dominated flow in the near field (i.e., close to the exit section) and a well-developed turbulent flow in the far field. In particular, \textit{Tokami et al.}~\cite{Tokami2020spatiotemporal} recently exploited the HVG approach for spatial series (instead of time-series) of velocity, suggesting that the near field order is related the gravitational contribute in the vorticity dynamics. Additionally, spatial network analyses of turbulent fires have also been carried out by employing the vorticity-induced approach of \textit{Taira et al.}~\cite{taira2016network}, under reference and augmented gravitational conditions~\cite{takagi2018synchronization, takagi2018effect}. Similarly to the 2D decaying isotropic turbulence (Section~\ref{subsec:isotrp}), the authors found a scale-free network structure, where the effects of gravity is to increase the slope of the (power-law) node-strength PDF~\cite{takagi2018effect}. Therefore, as in the case of 2D isotropic turbulence, the scale-free network structure can be exploited for control strategies by targeting network hubs. However, for turbulent fires the scale-free structure tend to intermittently appear and disappear and -- although this aspect need to be further clarified -- a preliminary interpretation was related to the ability of the complex network to adapt to variations in the vortical structure of the flow~\cite{takagi2018synchronization}.

		Finally, \textit{Murugesan et al.}~\cite{murugesan2019complex} employed $\epsilon$-recurrence and NVG approaches from time-series (see \tablename~\ref{tab:methds}) to study the synchronization of a low density jet subjected to an external periodic forcing. Network metrics (e.g., degree or the assortativity coefficient~\cite{newman2018networks}, see~\ref{app:nets}) showed significant variations in the proximity of critical amplitude modulation values, thus making complex networks a reliable tool to detect the onset of synchronization. Moreover, complex networks were able to distinguish between different routes to synchronization~\cite{murugesan2019complex}, emphasizing the network sensitivity to different dynamical processes.

		\subsection{Wakes and Vortical flows}\label{subsec:wakes}

			Wakes are an important category of fluid motion, as they appear in a large variety of flow applications. The interaction between a cylinder and the oscillatory motion in its wake -- which produces drag and lift forces on the cylinder -- was analyzed by \textit{Nair et al.}~\cite{nair2018networked}. The authors exploited a direct numerical simulation (DNS) of an unsteady wake to evaluate oscillatory motions via proper orthogonal decomposition (POD) of the velocity field. They built a network-based model to study the inter-relations and energy transfer between conjugate mode pairs, which represent the network nodes. Links were weighted by using linear regression on the POD fluctuation series extracted from each node~\cite{nair2018networked}. Even by perturbing mode energy and phase, the network model was able to capture the inter-dynamics between modes, and a feedback control strategy was developed to damp oscillations and reduce unsteady forces on the cylinder. More recently, in two preliminary investigations, \textit{Tao and Wu}~\cite{tao2019modified, wu2020evolution} extracted sub-structures of the velocity signals (identified in the network as three mutually linked nodes) through the HVG algorithm, and characterized such structures in terms of their intensity and temporal length (paying attention to large-scale structures~\cite{wu2020evolution}).
		
			A long-standing issue in studying high-dimensional systems such as turbulence concerns the development of reduced-order models able to capture the essential features of the flow. \textit{Gopalakrishnan Meena et al.}~\cite{meena2018network} proposed an approach that relies on community detection in vorticity-induced networks (see \figurename~\ref{fig:mets_eul}a), where communities -- extracted through modularity maximization~\cite{newman2018networks} -- identified coherent vortical structures. A reduced-order model was then obtained by grouping vortical elements belonging to the same community~\cite{meena2018network}. The authors firstly demonstrated the efficacy of the method for a discrete set of point vortices, and then extended the procedure to a DNS of the cylinder wake and a DNS of the wake flow over a NACA 0012 airfoil. By exploiting the reduced order-model, aerodynamic forces (i.e., lift and drag) were predicted with good accuracy with respect to DNS values, even in presence of noisy data. These results indicated that the community-based model captures the effect of the vortex shedding on the body-forces~\cite{meena2018network}.
			
			Reduced-order modeling via coherent structure identification has also been carried out in the Lagrangian perspective. A trajectory-similarity based approach (\figurename~\ref{fig:mets_lagr}b) called \textit{coherent structure coloring} was developed by \textit{Schlueter-Kuck and Dabiri}~\cite{schlueter2017coherent, schlueter2017identification} and \textit{Husic et al.}~\cite{husic2019simultaneous} to face the problem of coherent structure identification for sparse Lagrangian data. The advantages of this network-based approach rely on the lack of \textit{a priori} knowledge of the main features (such as number and size) of the structures present in the flow, and the robustness under data sparsification (which makes the approach an effective alternative to more typical tools as finite-time Lyapunov exponent~\cite{schlueter2017coherent}). The authors demonstrated the efficacy of the methodology for 2D vortical flows as Bickley jets (i.e., a simplified model for atmospheric zonal jets), quadruple-eddy flows (which serves as a simplified model for ocean drifters) and experimentally generated vortex rings. In this last configuration, coherent structure coloring was effectively applied to identify the regions of flow entrainment in a vortex ring, which is a flow that appears -- among others applications -- in medicine (e.g., in human heart), thus providing a technique to deal with sparse data coming, e.g., from non-invasive clinical methods~\cite{husic2019simultaneous}.

			The Lagrangian framework was also employed in~\cite{padberg2017network} to detect strong mixing zones by means of a trajectory-proximity-based criterion (\figurename~\ref{fig:mets_lagr}a). The main idea is that high mixing is present in zones where particle trajectories frequently come close in space (i.e., where fluid exchange is enhanced). Hence, by means of classical network metrics (e.g., degree centrality and clustering coefficient, see~\ref{app:nets}) the authors identified high and low mixing zones in a Bickley jet, and extended the method to experimental data from a stratospheric polar vortex~\cite{padberg2017network}. The results for these flow configurations indicate that, both on numerical and experimental data, the network metrics are able to highlight strong mixing regions.

			Finally, a promising application of complex network is represented by flow classification. With this aim, \textit{Krueger et al.}~\cite{krueger2019quantitative} employed a proximity-based spatial network approach based on Grabriel graph criterion (see \figurename~\ref{fig:mets_eul}b), providing -- in a robust and accurate way -- a classification of different vortical flows (such as von K{\'a}rm{\'a}n vortex streets) in presence of different kinds of flow perturbations. Network-based flow classifications can have positive implications in assessing similarity among different possible swimming or flying motions~\cite{krueger2019quantitative}.

		\subsection{Wall turbulence and convection}	
		
			Turbulent flows developing in presence of walls (e.g., channels or boundary layers) play a crucial role in industrial and natural processes, and they are strictly related to energy-saving issues. In a first attempt to characterize wall turbulence by means of complex networks, \textit{Liu et al.}~\cite{liu2010statistical} and \textit{Iacobello et al.}~\cite{iacobello2018visibility} constructed NVGs (\figurename~\ref{fig:mets_ts}c) on time-series of energy dissipation rate from a wind tunnel, and velocity from a DNS of a turbulent channel flow, respectively. The NVG approach was exploited to inherit important features of (wall) turbulence, such as the self-similarity of energy dissipation rate (emerging as a scale-free network)~\cite{liu2010statistical}, and the occurrence of peaks and small fluctuations in the signal~\cite{iacobello2018visibility}. In this last case, the authors provided preliminary insights into how the metric behaviors relate to the main turbulence features in the wall-normal direction~\cite{iacobello2018visibility}, making the NVG an alternative tool for distinguishing different dynamical flow regimes.
			
			DNSs of turbulent channel flows were also exploited by \textit{Iacobello et al.}~\cite{iacobello2018spatial, iacobello2019lagrangian} and \textit{Perrone et al.}~\cite{perronewall} to study the spatio-temporal dynamics of the velocity field in a Eulerian perspective~\cite{iacobello2018spatial} and turbulent mixing in a Lagrangian viewpoint~\cite{iacobello2019lagrangian, perronewall}. In the former case, a spatial correlation-network was built (e.g., see \figurename~\ref{fig:mets_eul}c), where links were activated if the correlation coefficient values between velocity signals exceeded a given threshold. The network analysis revealed the presence of long-range links -- called teleconnections -- between distant near-wall regions, representing kinematically similar regions at distant locations. Moreover, network hubs were found to cluster in elongated structures as an effect of mean flow advection. Physically, clusters of hub and their teleconnections were associated to the presence of coherent high- and low-speed (near wall) streaks~\cite{iacobello2018spatial}. 
			\\In the Lagrangian viewpoint~\cite{iacobello2019lagrangian, perronewall}, instead, turbulent mixing was studied by means of a temporal network, where nodes correspond to groups of fluid particles~\cite{iacobello2019lagrangian} or different discrete levels at varying wall-normal coordinates~\cite{perronewall}. Link weights depend on the number of connections (based on spatial proximity) between particles at each time~\cite{iacobello2019lagrangian}, or on the number of particles moving from one level to another level over time~\cite{perronewall}. The relative intensity of mean flow advection and wall-normal turbulent mixing on the particle dynamics was quantified, thus distinguishing between characteristic dispersion regimes~\cite{iacobello2019lagrangian} and highlighting the features of wall-normal mixing in the advection-mixing transient regime. Moreover, particle mutual distances were also accounted, thus providing a network-based formulation that generalizes other Lagrangian proximity-based approaches (e.g.,~\cite{padberg2017network}).

			It is worth mentioning the application of network-based tools for the identification of coherent structures in a low-dimensional model of an inhomogeneous shear flow, which comprises -- through nine modeling coefficients -- fundamental characteristics of wall turbulence~\cite{schmid2018description}. By combining a community detection approach and community clustering~\cite{newman2018networks}, coherent structures and intermittent events were identified as persistent and transients states associated with low and high transition probabilities, respectively~\cite{schmid2018description}.

		Flow structures have also been uncovered in a turbulent Rayleigh-B{\'e}nard convection between two horizontal plates, through a Lagrangian proximity-based network analysis~\cite{schneide2018probing}. For short-times, the authors identified Lagrangian network clusters that correspond to very large-scale structures (i.e., \textit{superstructures}) in the Eulerian framework of the turbulent convection flow~\cite{schneide2018probing}. The method is shown to work well for longer time scales than traditional approaches based on finite-time Lyapunov exponents. For long times, instead, the analysis revealed that trajectories that remain closely together over time were found to coincide with the center of circulation rolls.
		\\Finally, we mention the application of the HVG to study large-scale double-diffusive convection of a viscoelastic fluid in a porous medium~\cite{kondo2018chaotic}. The authors showed and quantified significant variations of the network characteristics for different porosity levels as a result of a flow destabilization arising from the increase in porosity.

		\subsection{Flows in turbulent combustors}\label{subsec:combust_flows}
		
		 	Turbulent combustors find applications in several engineering systems (e.g., rocket and gas-turbine engines), and are characterized by the interplay between combustion, turbulence and an acoustic field, eventually leading to a thermoacoustic instability that is a dangerous phenomenon for engines' life. The transition to thermoacoustic instability as well as its characterization (e.g., its high-dimensional dynamics~\cite{okuno2015dynamics}) have been the main object of network-based investigations.	In particular, several time-series analyses highlighted that the network-based approach is able to fully detect the different regimes of combustion dynamics (i.e., combustion noise, intermittency, combustion instability and thermoacoustic instability) in terms of different network structures for recurrence networks~\cite{godavarthi2017recurrence, kobayashi2018nonlinear}, ordinal partition networks~\cite{kobayashi2018nonlinear, kobayashi2019early, aoki2020dynamic}, and visibility graphs~\cite{murugesan2015combustion, murugesan2016detecting, guan2020intermittency}. Therefore, according to the specific features that the network exhibits, network metrics can be used as early-warning indicator of the characteristic regimes in the combustor.
			
			Besides the possibility to distinguish different dynamical regimes via time-series analysis, spatial networks have also been constructed to identify the most significant regions in the flow where a control strategy can be implemented. \textit{Krishnan et al.}~\cite{krishnan2019emergence} used a spatial proximity approach to build (time-varying) networks from local acoustic power and vorticity fields. Following the temporal evolution of network metrics (e.g., link density), the authors found out large areas of acoustic power production during thermoacoustic instability while small clusters appear during combustion noise. \textit{Unni et al.}~\cite{unni2018emergence} built correlation networks from the velocity field to detect critical spatial regions in a combustor with a flame-holding device, finding critical regions under thermoacoustic instability as zones with very high values of network metrics such as the degree. Successively, \textit{Krishnan et al.}~\cite{krishnan2019mitigation} found out that critical regions identified via correlation spatial networks are indeed the optimal spatial locations for passive control of oscillatory instabilities. 
			\\Vorticity-induced (weighted) spatial networks were also used to detect thermoacoustic combustion oscillations~\cite{murayama2018characterization, hashimoto2019spatiotemporal}. Scale-free network structures appeared, in which hubs are associated to minimum of pressure fluctuations near the injector~\cite{murayama2018characterization} and hubs periodically appear with a characteristic acoustic-mode period~\cite{hashimoto2019spatiotemporal}. Therefore, network hubs in vorticity networks, due to the scale-free property, can be targeted (as for isotropic turbulence) for the implementation of control strategies.

	Finally, differently from the usual time-series analysis, \textit{Singh et al.}~\cite{singh2017network} used the visibility criterion to capture the spatial structure of flame fronts (i.e., the interface between reactants and products). Nodes with high visibility, representing the node hubs, were found in large-curvature folded regions of the flame front, which are associated to the effect of turbulent eddies on the flame front.
	
		\begin{figure}[t]
			\centering
			\includegraphics[width=\textwidth]{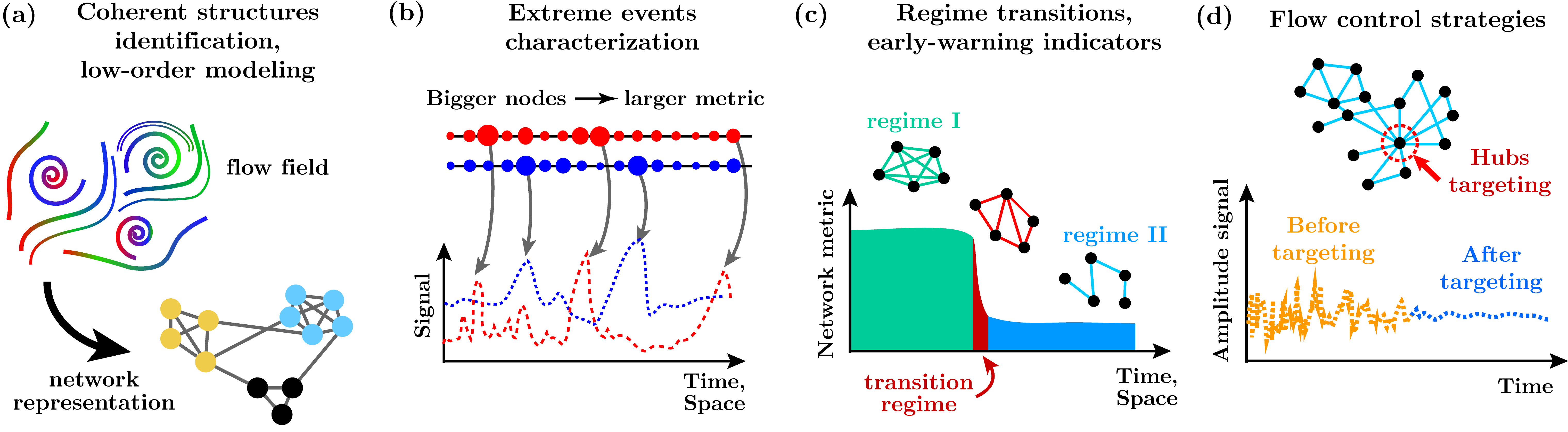}
			\caption{Schematic of the main achievements gained in turbulent and vortical flows via complex networks. (a) Coherent structure identification and reduced-order modeling: a sketch of a flow field with three vortices, and the corresponding network where different node colors highlight the three main vortical structures. (b) Extreme events characterization, showing (bottom) two signals and (top) the corresponding networks (only nodes are shown), where larger node sizes identify signal peaks. (c) Dynamics transition detection (in time or space) showing the variation of a network metric for different regimes (highlighted as green, red and blue regions). (d) An example of a control strategy, showing the change of a characteristic flow variable before and after a control implementation based on network hubs targeting.} \label{fig:res_schem}
		\end{figure}

	\section{Discussion and conclusions}\label{sec:concl}

		Complex networks have shown a high effectiveness in capturing key aspects of the dynamics of different turbulent flows, and several reviewed works emphasized the ability of networks to represent a valid tool for the analysis of turbulent flows. The key feature emerging is the extreme versatility of complex networks. They have been employed for studying several flow configurations, each one with its peculiarities. For example, jets and plumes (Section~\ref{subsec:jets}) have been focused under heating and cooling conditions or in presence of buoyancy effects, different arrangements of vortical flows have been studied (Section~\ref{subsec:combust_flows}), and turbulent combustors with different flame stabilizers have been considered (Section~\ref{subsec:wakes}).
		
		 Complex networks theory, hence, provides a wide spectrum of tools to map and analyze the spatio-temporal data coming from numerical simulations or experiments. In particular, the results obtained from numerical and experimental data are shown to be rather consistent (e.g., see~\cite{charakopoulos2014application, padberg2017network, husic2019simultaneous, kobayashi2019spatiotemporal}, although experimental results are more noisy. From the point of view of data transformation, a geometrical representation can be implemented through a network technique that best suits the data availability (as outlined in \tablename~\ref{tab:methds}), whether if data come from spatial or temporal (univariate or multivariate) signals, multidimensional spatio-temporal fields (as those from DNSs), or as Lagrangian trajectories. Furthermore, the network analysis can be effectively combined with other techniques to increase the level of information on the flow system, even by defining \textit{ad hoc} metrics. Examples are the construction of networks from POD modes for studying oscillatory motions~\cite{nair2018networked}, or by pre-projecting the data into a properly-defined phase-space, and building networks from the information conveyed to such phase-spaces. The possibility to integrate the network approaches with existing techniques lends an additional level of flexibility to complex networks.
		 
		 This methodological versatility translates into a variety of ways to extract informative features of the dynamics of turbulence and other vortical flows. Although the use of complex network-based approaches in turbulence is very recent and there is not yet enough material to draw quantitative conclusions, it is nevertheless useful to highlight the main results achieved so far, as shown in \figurename~\ref{fig:res_schem}.	Specifically, as reviewed in the present work, networks have been used to identify flow features such as Eulerian or Lagrangian coherent structures (\figurename~\ref{fig:res_schem}a), to characterize the appearance of critical spatial regions and temporal extreme events (\figurename~\ref{fig:res_schem}b), as well as to distinguish between different dynamical regimes (\figurename~\ref{fig:res_schem}c). The detection of critical regions in the flow -- e.g., the formation of coherent motion -- promoted the implementation of effective control strategies (\figurename~\ref{fig:res_schem}d) and the definition of lower-order models. These two features are crucial in fluid dynamics, since flows usually exhibit a high-level of complexity that, in several contexts, need to be properly controlled. Besides, the characterization of extreme events and the transition between different regimes have important practical implications (e.g., for dispersion processes or unstable-to-transitional regimes). From this perspective, network metrics revealed to be reliable early-warning indicators, and powerful tools in capturing the appearance and intensity of extreme events. Eventually, flow reconstruction from the network topology -- although at present there are only very few proposals -- can be a promising application in the future. Therefore, complex networks show a remarkable capability to capture spatial and temporal features of turbulent and vortical flows, in a different way with respect to other techniques.

		In the face of these qualities, however, some delicate aspects should be mentioned. If, on one hand, the wide range of techniques and metrics provides versatility, on the other hand it could also be a drawback, at least at the early stages of analysis when two questions usually emerge: (i) which is the most suitable approach for mapping a given dataset in a network?, and (ii) which are the most informative network metrics to shed light on a given flow-related issue? The answers to these questions are not trivial and depend, among others, on the flow configuration and the aim of the work, the data availability and the computational resources. Concerning the choice of a suitable network-based method to study a given type of turbulent or vortical dataset (i.e., signals, multidimensional fields or trajectories), still there is not a simple and unique \textit{vade mecum} that can guide one in identifying the best network approach for attaining the aim of a turbulence analysis. The network-based applications reported so far have provided preliminary insights on what a given network approach can point out for a specific flow configuration. For example, visibility graphs have been effectively employed to characterize the intensity of extreme events, while recurrence networks reveal to be more informative on the recursive appearance of events. However, a general perspective still lacks. 
		\\At the same time, the choice of a network metric that highlights a specific flow feature, requires an understanding of the physical meaning that metric has in the context in which is used. This aspect is probably one of the hardest tasks in building bridges between network science and turbulence, because it calls for great endeavors to physically interpret the network metrics. Otherwise, one only gets a quantitative description of the metrics, without a physical interpretation of it. Moreover, at the current stage, there are not sufficient elements to provide reliable discussions on critical issues such as universality or uniqueness of the network-based methods. These two topics -- i.e., the physical interpretation of network metrics and   a discussion on critical methodological issues of networks -- surely deserve further investigation in future analyses.
		
		The quality of the data can also affect the network results. In general, a good spatial or temporal resolution is required to capture sufficient information from the flow. In Lagrangian network approaches, the temporal window, the time-step and the total number of particles seeded in the flow need to be chosen in order to capture the phenomenon under study. A low number of particles may not be able to capture the flow dynamics while an elevated number of particles lead to large networks (i.e., high computational costs). Due to the need of a sufficiently accurate spatial resolution to characterize the smallest significant scales, spatial networks (especially from 3D data) usually result in very large graphs, i.e. networks with a huge number of nodes and links. To overcome dimensionality reduction issues, different approaches have been pursued, such as by relying on a sparsification technique based on graph spectral properties~\cite{nair2015network}, on network community-based approaches~\cite{meena2018network}, on matrix randomization techniques~\cite{bai2019randomized}, as well as by properly tuning the threshold levels in correlation networks~\cite{scarsoglio2016complex, iacobello2018spatial}. Several optimized algorithms have also been proposed for building visibility graphs, deeply reducing their computational costs~\cite{lan2015fast, yela2020online}. It should be noted that, the invariance of visibility graphs to affine transformations of the series can be a drawback if the analysis requires a method that is sensitive to rescaling of the axes (e.g., the series mean value), while the invariance is a potential benefit if the focus is on the temporal structure of the signals. In addition, similarly to other signal analysis tools, two characteristics affect to some extent all the three network-based methods, i.e., the sampling frequency and the series length (e.g., see~\cite{donner2012visibility} for visibility graphs). Therefore, the features of the setup (e.g., number of particles, spatial and temporal resolution, domain size, etc) have to properly be accounted for, but the network building should only rely on the specific physical insights looked for in the flow of interest.

		In conclusion, although the application of complex networks in turbulence studies is still in its early stages and despite the intrinsic limitations that one may encounter, positive hints have emerged from the growing number of applications and analyses. Alongside the developments of novel insights into complex network theory~\cite{newman2018networks, barabasi2012network}, we believe that complex networks can soon be established as a widespread tool for turbulence analysis, so that more quantitative comparisons and assessments can be done once the topic of complex networks in fluid dynamics will reach a sufficiently-grown stage. In this vein, we hope this review can boost future network-based research on turbulent and vortical flows.

\appendix
	\section{A methodological introduction to complex networks and some their metrics}\label{app:nets}

		In this section we provide a synthetic overview about complex networks, including some main definitions of frequently-used network metrics. In particular, this appendix is intended to give a brief methodological background that would support a reader who is not accustomed to the network formalism. The reader is referred to~\cite{boccaletti2006complex, costa2007characterization, holme2012temporal, newman2018networks} for comprehensive and detailed discussions on complex networks and their metrics.

		Complex networks arise as a tool to geometrically represent the mutual inter-connections among the elements of a real-world system. From a mathematical viewpoint, a network is defined as a graph $G(N_v,N_e) = (\mathcal{V},\mathcal{E})$, where $\mathcal{V}=\lbrace 1,2,...,N_v\rbrace$ is a set of $N_v$ labeled nodes (or vertices) and $\mathcal{E}=\lbrace 1,2,...,N_e\rbrace$ is a set of $N_e$ links (or edges)~\cite{newman2018networks}. The inter-connections between nodes are usually represented by means of a binary and symmetric adjacency matrix, $A_{i,j}$, defined as
		
		\begin{equation}
			A_{i,j} = \left \{\begin{array}{l} 0$, if $i=j$ or $\lbrace i,j\rbrace\not\in\mathcal{E},\\ 1$, if $\lbrace i,j\rbrace\in\mathcal{E}.
			\end{array}	
			\right.
		\end{equation}

		In other words, $A_{i,j}=1$ if nodes $i$ and $j$ are linked and $0$ otherwise. This is the simplest network formulation as the presence or absence of a connection between two nodes is retained in $A_{i,j}$. More general formulations include the possibility to account for node or link intensity (i.e., by assigning a weight to each node and/or each link) leading to weighted networks, to retain the link direction (i.e., $A_{i,j}\neq A_{j,i}$) leading to directed networks, as well as the possibility to account for higher-order interactions (for a comprehensive review see~\cite{battiston2020networks}). Moreover, both static and time-varying networks can be built, namely networks in which the features of nodes and/or links remain constant or change in time, respectively~\cite{holme2012temporal}.
		
		The versatility of complex networks in representing discrete systems by means of a spectrum of different formulations has also led to an extremely rapid development of tools for network analysis. Here we recall the definition of some network metrics that are widely employed in network science, specifically in the works reviewed in Section~\ref{sec:applications}. Network metrics can be distinguished into local and global metrics, corresponding to two different scales at which the metrics are evaluated. Local metrics are associated to single nodes, while global metrics refer to the entire network. In what follows, we recall the definition and properties of a few network metrics, including three principal categories: centrality, clustering and assortativity metrics~\cite{costa2007characterization, newman2018networks, boccaletti2006complex}.

		\subsection{Centrality indices}
			
			Centrality metrics are introduced to quantify the relative importance of nodes in the network. The degree centrality represents the most simple centrality metric, and is defined as the number of links connected to a node $i$. In case of weighted network formulations, $A_{i,j}$ entries represent scalar values so that the weighted-counterpart of the degree is referred to as strength~\cite{newman2018networks}. The nodes linked to $i$ are its neighbors, so the degree gives the cardinality of the neighborhood of $i$. 
			\\An important statistical measure associated to the degree (or the strength) is its degree probability distribution, namely the probability to find nodes with given values of degree (or strength). In fact, the degree distribution is used to discriminated between different classes of networks, such as random or scale-free networks~\cite{newman2018networks}. For instance, power-law degree distributions have been associated with network robustness under random perturbations~\cite{taira2016network}. Additionally, a widely employed metric in turbulent and vortical flow analysis is the network entropy, which is the Shannon entropy of the degree distribution and is used to quantify the randomness in the studied complex system (e.g., see~\cite{kobayashi2019spatiotemporal}).
			
			Further widely employed centrality metrics include eigenvector centrality, the betweenness centrality and the closeness centrality~\cite{newman2018networks}. 
			\\The eigenvector centrality of a node $i$ is an extension of the degree centrality, as it takes into account the centrality values of the neighbors of $i$ to quantify the importance of $i$~\cite{newman2018networks}. 
			\\The betweenness and the closeness centrality, instead, are based on the concept of shortest path, namely the shortest alternating sequence of nodes and links (considered only once) to reach a node $j$ starting from another node $i$, and \textit{vice versa}~\cite{boccaletti2006complex}. High betweenness centrality values correspond to nodes involved in several shortest paths, while the closeness centrality of a node $i$ indicates the average length of the shortest paths from $i$ to all the other nodes in the network.
			
			Centrality metrics are usually referred to single nodes, so that global centrality metrics are typically obtained by averaging the corresponding local metrics over all network nodes. However, it is worth to mention here two global metrics, the average path length and the network density, that are also widely used in network analysis. The average path length is defined as the average of all shortest path lengths in a network, and its importance relies on the fact that it provides an average measure of the efficiency to reach a given node starting from another node in the network~\cite{boccaletti2006complex}. 
			\\The network density, instead, is quantified as the ratio between the number of links, $N_e$, and the total number of possible links in the network~\cite{newman2018networks}. The network density thus gives a measure of the sparsity of the a graph.

		\subsection{Clustering metrics}
			
			An important feature of many networks is the presence of three-nodes relationships, which are quantified by clustering metrics. A local clustering metric is the so called Watts-Strogatz clustering coefficient, that quantifies the probability that two randomly chosen nodes, $j$ and $h$, both linked to a third node $i$ (i.e., $A_{i,j}=A_{i,h}=1$) are also connected (i.e., $A_{j,h}=1$)~\cite{boccaletti2006complex}. A global clustering coefficient can be evaluated by averaging its local counterpart over all nodes. 
			\\Another clustering metric is the transitivity, that is a global metric. It is defined as the fraction between the number of triangles and the number of triples in the network, where a triangle is a set of three nodes all linked between them	while a triple is a set of three nodes in which (at least) two of them are directly linked to the third node~\cite{costa2007characterization}.
			
			Clustering metrics are employed to investigate the extent to which the neighborhood of a node is interconnected, thus representing a measure of the presence of an ordered behavior in the system. Real-world systems typically show high levels of three-nodes relationships, so high clustering coefficients are usually obtained.

		\subsection{Assortativity metrics}
			
			While centrality metrics are useful to characterize the network structure, they do not directly quantify whether the centrality of a node is similar or not to the centrality of other nodes. Assortativity metrics are introduced to face this issue, so a network is said to be assortative or disassortative if nodes tend
to link with other similar or dissimilar nodes, respectively~\cite{newman2018networks}. Similarity is typically measured through a centrality metric (such as the degree or the strength). If nodes do not tend to link neither to similar nor dissimilar other nodes, the network is said to be non-assortative.
			\\The most widely used measure of assortativity is the assortativity coefficient, which is a global metric defined as the Pearson correlation coefficient of the degree of the nodes at the ends of each link. In fact, the sequence of links $\lbrace i,j\rbrace$ in the network, with $i,j=1,\dots,N_v$ and $A_{i,j}=1$, form two vectors of nodes (one for $i$ and the other for the $j$ index) where each node has its own degree centrality value. In this way, two vectors are obtained whose elements are the degree centrality values for each pair of linked nodes. The assortativity coefficient is then the (Pearson) correlation coefficient between these two degree vectors, thus ranging in the interval $\left[-1,1\right]$. Assortative, disassortative and non-assortative networks are characterized by positive, negative and null values of the assortativity coefficient, respectively.

		\subsection{Customized metrics}
		
		The three categories of network metrics discussed above provide the most basic tools to shed light on the network structure, and have been widely employed in network-based investigations of turbulent and vortical flows. However, thanks to the versatility of complex networks, other metrics have also been defined \textit{ad hoc} to accommodate the various needs of interpretation of results arising in different studies. An example is the mean link length, which was defined in~\cite{iacobello2018visibility} as a local metric to quantify the average temporal distance between a node and its neighborhood in the context of visibility graphs. The average peak occurrence was also defined as the inverse of the mean link length~\cite{iacobello2019experimental}, to highlight low values of the metric. Another distance-based metric was also proposed in~\cite{scarsoglio2016complex, iacobello2018spatial} to quantify the average (Euclidean) distance between nodes in spatial networks. 
		\\Additionally, \textit{Taira et al.}~\cite{taira2016network} proposed a normalized variant of the average path length, with the aim to characterize the network changes under node perturbations. In fact, the perturbation or removal of nodes from the network significantly affect the shortest path lengths, whose average value can be used to assess the resilience of the network.

\biboptions{sort&compress}
\bibliography{biblio_TurbNets_Rev}

\begin{thebibliography}{10}

\bibitem{pope2001turbulent}
Stephen~B Pope.
\newblock {\em Turbulent flows}.
\newblock Cambridge University Press, (2001).

\bibitem{wu2015vortical}
J.Z. Wu, H.Y. Ma, and M.D. Zhou.
\newblock {\em Vortical flows}.
\newblock Springer, (2015).

\bibitem{klewicki2010reynolds}
J.~C. Klewicki.
\newblock {\em J. Fluid Eng.}, 132(9), 2010.

\bibitem{marusic2010wall}
I.~Marusic, B.J. McKeon, P.A. Monkewitz, H.M. Nagib, A.J. Smits, and K.R.
  Sreenivasan.
\newblock {\em Phys. Fluids}, 22(6):065103, 2010.

\bibitem{sujith2020complex}
R.I. Sujith and V.~R. Unni.
\newblock {\em Phys. Fluids}, 32(6):061401, 2020.

\bibitem{pollard2017whither}
A.~Pollard, L.~Castillo, L.~Danaila, and M.~Glauser.
\newblock In {\em Whither Turbulence and Big Data in the 21st Century?}, pages
  531--547. Springer, 2017.

\bibitem{haller2001distinguished}
G.~Haller.
\newblock {\em Physica D}, 149(4):248--277, 2001.

\bibitem{brunton2020machine}
S.~L. Brunton, B.~R. Noack, and P.~Koumoutsakos.
\newblock {\em Annu. Rev. Fluid Mech.}, 52:477--508, 2020.

\bibitem{costa2011analyzing}
L.~da~F.~Costa, O.~N. Oliveira~Jr, G.~Travieso, F.~A. Rodrigues, P.~R.
  Villas~Boas, L.~Antiqueira, M.~P. Viana, and L.~E. Correa~Rocha.
\newblock {\em Adv. Phys.}, 60(3):329--412, 2011.

\bibitem{donner2017focusissue}
R.V. Donner, E.~Hern{\'{a}}ndez-García, and E.~Ser-Giacomi.
\newblock {\em Chaos}, 27(3):035601, 2017.

\bibitem{nocke2015visual}
T.~Nocke, S.~Buschmann, J.~F. Donges, N.~Marwan, H.J. Schulz, and C.~Tominski.
\newblock {\em Nonlinear Proc. Geoph.}, 22(5):545, 2015.

\bibitem{dijkstra2019networks}
H.~A. Dijkstra, E.~Hern{\'a}ndez-Garc{\'\i}a, C.~Masoller, and M.~Barreiro.
\newblock {\em Networks in Climate}.
\newblock Cambridge University Press, 2019.

\bibitem{agarwal2019network}
A.~Agarwal, L.~Caesar, N.~Marwan, R.~Maheswaran, B.~Merz, and J.~Kurths.
\newblock {\em Sci. Rep.}, 9(1):1--12, 2019.

\bibitem{ying2020rossby}
N.~Ying, D.~Zhou, Z.G. Han, Q.H. Chen, Q.~Ye, and Z.G. Xue.
\newblock {\em Geophys. Res. Lett.}, 47(2):e2019GL086507, 2020.

\bibitem{zou2019complex}
Y.~Zou, R.~V. Donner, N.~Marwan, J.~F. Donges, and J.~Kurths.
\newblock {\em Phys. Rep.}, 787:1--97, 2019.

\bibitem{marwan2009complex}
N.~Marwan, J.~F. Donges, Y.~Zou, R.~V. Donner, and J.~Kurths.
\newblock {\em Phys. Lett. A}, 373(46):4246--4254, 2009.

\bibitem{zhang2006complex}
Jie Zhang and Michael Small.
\newblock {\em Phys. Rev. Lett.}, 96(23):238701, 2006.

\bibitem{takagi2017nonlinear}
Kazushi Takagi, Hiroshi Gotoda, Isao~T Tokuda, and Takaya Miyano.
\newblock {\em Phys. Rev. E}, 96(5):052223, 2017.

\bibitem{kobayashi2018nonlinear}
H.~Kobayashi, H.~Gotoda, and S.~Tachibana.
\newblock {\em Physica A}, 510:345--354, 2018.

\bibitem{godavarthi2017recurrence}
V.~Godavarthi, V.R. Unni, E.A. Gopalakrishnan, and R.I. Sujith.
\newblock {\em Chaos}, 27(6):063113, 2017.

\bibitem{murugesan2019complex}
M.~Murugesan, Y.~Zhu, and L.~K.B. Li.
\newblock {\em Int. J. Heat Fluid F.}, 76:14--25, 2019.

\bibitem{gotoda2017characterization}
H.~Gotoda, H.~Kinugawa, R.~Tsujimoto, S.~Domen, and Y.~Okuno.
\newblock {\em Phys. Rev. Appl.}, 7(4):044027, 2017.

\bibitem{charakopoulos2014application}
A.K. Charakopoulos, T.E. Karakasidis, P.N. Papanicolaou, and A.~Liakopoulos.
\newblock {\em Chaos}, 24(2):024408, 2014.

\bibitem{okuno2015dynamics}
Y.~Okuno, M.~Small, and H.~Gotoda.
\newblock {\em Chaos}, 25(4):043107, 2015.

\bibitem{xu2008superfamily}
X.~Xu, J.~Zhang, and M.~Small.
\newblock {\em PNAS}, 105(50):19601--19605, 2008.

\bibitem{Kaiser2014Cluster}
E.~Kaiser, B.~R. Noack, L.~Cordier, A.~Spohn, M.~Segond, M.~Abel, G.~Daviller,
  J.~Östh, S.~Krajnović, and R.~K. Niven.
\newblock {\em J. Fluid Mech.}, 754:365–414, 2014.

\bibitem{nair2019cluster}
A.~G. Nair, C.A. Yeh, E.~Kaiser, B.~R. Noack, S.~L. Brunton, and K.~Taira.
\newblock {\em J. Fluid Mech.}, 875:345--375, 2019.

\bibitem{newman2018networks}
M.~Newman.
\newblock {\em Networks}.
\newblock Oxford University Press, 2 edition, (2018).

\bibitem{mccullough2015time}
M.~McCullough, M.~Small, T.~Stemler, and H.~H. Iu.
\newblock {\em Chaos}, 25(5):053101, 2015.

\bibitem{lacasa2008time}
L.~Lacasa, B.~Luque, F.~Ballesteros, J.~Luque, and J.~C. Nuno.
\newblock {\em PNAS}, 105(13):4972--4975, 2008.

\bibitem{luque2009horizontal}
B.~Luque, L.~Lacasa, F.~Ballesteros, and J.~Luque.
\newblock {\em Phys. Rev. E}, 80(4):046103, 2009.

\bibitem{iacobello2019experimental}
G.~Iacobello, M.~Marro, L.~Ridolfi, P.~Salizzoni, and S.~Scarsoglio.
\newblock {\em Phys. Rev. Fluids}, 4(10):104501, 2019.

\bibitem{nair2015network}
A.~G. Nair and K.~Taira.
\newblock {\em J. Fluid Mech.}, 768:549--571, 2015.

\bibitem{taira2016network}
K.~Taira, A.G. Nair, and S.L. Brunton.
\newblock {\em J. Fluid Mech.}, 795:R2, 2016.

\bibitem{krueger2019quantitative}
P.~S. Krueger, M.~Hahsler, E.~V. Olinick, S.~H. Williams, and M.~Zharfa.
\newblock {\em Proc. R. Soc. A}, 475(2228):20180897, 2019.

\bibitem{aste2006dynamical}
T.~Aste and T.~Di~Matteo.
\newblock {\em Physica A}, 370(1):156--161, 2006.

\bibitem{donges2009complex}
J.F. Donges, Y.~Zou, N.~Marwan, and J.~Kurths.
\newblock {\em Eur. Phys. J. - Spec. Top.}, 174(1):157--179, 2009.

\bibitem{zhou2015teleconnection}
D.~Zhou, A.~Gozolchiani, Y.~Ashkenazy, and S.~Havlin.
\newblock {\em Phys. Rev. Lett.}, 115(26):268501, 2015.

\bibitem{tupikina2016correlation}
L.~Tupikina, N.~Molkenthin, C.~L{\'o}pez, E.~Hern{\'a}ndez-Garc{\'\i}a,
  N.~Marwan, and J.~Kurths.
\newblock {\em PloS One}, 11(4):e0153703, 2016.

\bibitem{meena2018network}
M.~Gopalakrishnan~Meena, A.~G. Nair, and K.~Taira.
\newblock {\em Phys. Rev. E}, 97(6):063103, 2018.

\bibitem{hadjighasem2016spectral}
A.~Hadjighasem, D.~Karrasch, H.~Teramoto, and G.~Haller.
\newblock {\em Phys. Rev. E}, 93(6):063107, 2016.

\bibitem{padberg2017network}
K.~Padberg-Gehle and C.~Schneide.
\newblock {\em Nonlinear Proc. Geoph.}, 24(4):661, 2017.

\bibitem{schneide2018probing}
C.~Schneide, A.~Pandey, K.~Padberg-Gehle, and J.~Schumacher.
\newblock {\em Phys. Rev. Fluids}, 3(11):113501, 2018.

\bibitem{iacobello2019lagrangian}
G.~Iacobello, S.~Scarsoglio, J.G.M. Kuerten, and L.~Ridolfi.
\newblock {\em J. Fluid Mech.}, 865:546--562, 2019.

\bibitem{schlueter2017coherent}
K.~L. Schlueter-Kuck and J.~O. Dabiri.
\newblock {\em J. Fluid Mech.}, 811:468--486, 2017.

\bibitem{husic2019simultaneous}
B.~E. Husic, K.~L. Schlueter-Kuck, and J.~O. Dabiri.
\newblock {\em PloS One}, 14(3):e0212442, 2019.

\bibitem{ser2015flow}
E.~Ser-Giacomi, V.~Rossi, C.~L{\'o}pez, and E.~Hern{\'a}ndez-Garc{\'\i}a.
\newblock {\em Chaos}, 25(3):036404, 2015.

\bibitem{shirazi2009mapping}
A.H. Shirazi, G.~R. Jafari, J.~Davoudi, J.~Peinke, M.~R.~R. Tabar, and
  M.~Sahimi.
\newblock {\em J. Stat. Mech.}, 2009(07):P07046, 2009.

\bibitem{schmid2018description}
P.J. Schmid, A.~Garc{\'\i}a-Gutierrez, and J.~Jim{\'e}nez.
\newblock In {\em J. Phys. Conf. Ser.}, volume 1001, page 012015. IOP
  Publishing, (2018).

\bibitem{kobayashi2019early}
T.~Kobayashi, S.~Murayama, T.~Hachijo, and H.~Gotoda.
\newblock {\em Phys. Rev. Appl.}, 11(6):064034, 2019.

\bibitem{kobayashi2019spatiotemporal}
W.~Kobayashi, H.~Gotoda, S.~Kandani, Y.~Ohmichi, and S.~Matsuyama.
\newblock {\em Chaos}, 29(12):123110, 2019.

\bibitem{aoki2020dynamic}
C.~Aoki, H.~Gotoda, S.~Yoshida, and S.~Tachibana.
\newblock {\em J. Appl. Phys.}, 127(22):224903, 2020.

\bibitem{iacobello2018complex}
G.~Iacobello, L.~Ridolfi, M.~Marro, P.~Salizzoni, and S.~Scarsoglio.
\newblock In {\em Progress in Turbulence VIII}, pages 215--220. Springer,
  (2019).

\bibitem{manshour2015fully}
Pouya Manshour, M~Reza~Rahimi Tabar, and Joachim Peinke.
\newblock {\em J. Stat. Mech.}, 2015(8):P08031, 2015.

\bibitem{takagi2018dynamic}
K.~Takagi, H.~Gotoda, I.~T. Tokuda, and T.~Miyano.
\newblock {\em Phys. Lett. A}, 382(44):3181--3186, 2018.

\bibitem{Tokami2020spatiotemporal}
T.~Tokami, T.~Hachijo, T.~Miyano, and H.~Gotoda.
\newblock {\em Phys. Rev. E}, 101:042214, 2020.

\bibitem{tao2019modified}
X.~Tao and H.~Wu.
\newblock In {\em AIAA Scitech Forum}, page 1868, (2019).

\bibitem{wu2020evolution}
H.~Wu and X.~Tao.
\newblock In {\em AIAA Scitech Forum}, page 0353, (2020).

\bibitem{liu2010statistical}
C.~Liu, W.~Zhou, and W.~Yuan.
\newblock {\em Physica A}, 389(13):2675--2681, 2010.

\bibitem{iacobello2018visibility}
G.~Iacobello, S.~Scarsoglio, and L.~Ridolfi.
\newblock {\em Phys. Lett. A}, 382(1):1--11, 2018.

\bibitem{kondo2018chaotic}
S.~Kondo, H.~Gotoda, T.~Miyano, and I.~T. Tokuda.
\newblock {\em Physica D}, 364:1--7, 2018.

\bibitem{murugesan2015combustion}
Me. Murugesan and R.I. Sujith.
\newblock {\em J. Fluid Mech.}, 772:225--245, 2015.

\bibitem{murugesan2016detecting}
M.~Murugesan and R.I. Sujith.
\newblock {\em J. Propul. Power}, 32(1):707--712, 2016.

\bibitem{guan2020intermittency}
Y.~Guan, V.~Gupta, and L.K.B. Li.
\newblock {\em J. Fluid Mech.}, 894, 2020.

\bibitem{singh2017network}
J.~Singh, R.~Belur~Vishwanath, S.~Chaudhuri, and R.I. Sujith.
\newblock {\em Chaos}, 27(4):043107, 2017.

\bibitem{bai2019randomized}
Z.~Bai, N.B. Erichson, M.~Gopalakrishnan~Meena, K.~Taira, and S.L. Brunton.
\newblock {\em PloS One}, 14(11), 2019.

\bibitem{takagi2018synchronization}
K.~Takagi, H.~Gotoda, T.~Miyano, S.~Murayama, and I.~T. Tokuda.
\newblock {\em Chaos}, 28(4):045116, 2018.

\bibitem{takagi2018effect}
K.~Takagi and H.~Gotoda.
\newblock {\em Phys. Rev. E}, 98(3):032207, 2018.

\bibitem{murayama2018characterization}
S.~Murayama, H.~Kinugawa, I.~T. Tokuda, and H.~Gotoda.
\newblock {\em Phys. Rev. E}, 97(2):022223, 2018.

\bibitem{hashimoto2019spatiotemporal}
T.~Hashimoto, H.~Shibuya, H.~Gotoda, Y.~Ohmichi, and S.~Matsuyama.
\newblock {\em Phys. Rev. E}, 99(3):032208, 2019.

\bibitem{krishnan2019emergence}
A.~Krishnan, R.I. Sujith, N.~Marwan, and J.~Kurths.
\newblock {\em J. Fluid Mech.}, 874:455--482, 2019.

\bibitem{scarsoglio2016complex}
S.~Scarsoglio, G.~Iacobello, and L.~Ridolfi.
\newblock {\em Int. J. Bifurcat. Chaos}, 26(13):1650223, 2016.

\bibitem{iacobello2018spatial}
G.~Iacobello, S.~Scarsoglio, J.G.M. Kuerten, and L.~Ridolfi.
\newblock {\em Phys. Rev. E}, 98(1):013107, 2018.

\bibitem{unni2018emergence}
V.~R. Unni, A.~Krishnan, R.~Manikandan, N.~B. George, R.I. Sujith, N.~Marwan,
  and J.~Kurths.
\newblock {\em Chaos}, 28(6):063125, 2018.

\bibitem{krishnan2019mitigation}
A.~Krishnan, R.~Manikandan, P.R. Midhun, K.V. Reeja, V.R. Unni, R.I. Sujith,
  N.~Marwan, and J.~Kurths.
\newblock {\em EPL}, 128(1):14003, 2019.

\bibitem{schlueter2017identification}
K.~L. Schlueter-Kuck and J.~O. Dabiri.
\newblock {\em Chaos}, 27(9):091101, 2017.

\bibitem{perronewall}
D.~Perrone, J.~G.~M. Kuerten, L.~Ridolfi, and S.~Scarsoglio.
\newblock {\em Phys. Rev. E}, 102(4), 2020.

\bibitem{nair2018networked}
A.~G. Nair, S.~L. Brunton, and K.~Taira.
\newblock {\em Phys. Rev. E}, 97(6):063107, 2018.

\bibitem{lan2015fast}
X.~Lan, H.~Mo, S.~Chen, Q.~Liu, and Y.~Deng.
\newblock {\em Chaos}, 25(8):083105, 2015.

\bibitem{yela2020online}
D.~F. Yela, F.~Thalmann, V.~Nicosia, D.~Stowell, and M.~Sandler.
\newblock {\em Phys. Rev. Res.}, 2(2):023069, 2020.

\bibitem{donner2012visibility}
R.~V. Donner and J.~F. Donges.
\newblock {\em Acta Geophys.}, 60(3):589--623, 2012.

\bibitem{barabasi2012network}
A.L. Barab{\'a}si.
\newblock {\em Nature Physics}, 8(1):14--16, 2012.

\bibitem{boccaletti2006complex}
S.~Boccaletti, V.~Latora, Y.~Moreno, M.~Chavez, and D.U. Hwang.
\newblock {\em Phys. Rep.}, 424(4):175--308, 2006.

\bibitem{costa2007characterization}
L.F. Costa, F.A Rodrigues, G.~Travieso, and P.R. Villas~Boas.
\newblock {\em Adv. Phys.}, 56(1):167--242, 2007.

\bibitem{holme2012temporal}
P.~Holme and J.~Saram{\"a}ki.
\newblock {\em Phys. Rep.}, 519(3):97--125, 2012.

\bibitem{battiston2020networks}
F.~Battiston, G.~Cencetti, I.~Iacopini, V.~Latora, M.~Lucas, A.~Patania, J.-G.
  Young, and G.~Petri.
\newblock {\em Phys. Rep.}, 874:1--92, 2020.

\end{thebibliography}

\end{document}